# Solar storms and submarine internet cables

Jorge C. Castellanos, Jo Conroy, Valey Kamalov, Mattia Cantono, and Urs Hölzle

Google LLC, Mountain View, CA, USA.

## Abstract

Coronal mass ejections (CMEs) can trigger geomagnetic storms and induce geoelectric currents that degrade the performance of terrestrial power grid operations; in particular, CMEs are known for causing large-scale outages in electrical grids. Submarine internet cables are powered through copper conductors spanning thousands of kilometers and are vulnerable to damage from CMEs, raising the possibility of a large-scale and long-lived internet outage. To better understand the magnitude of these risks, we monitor voltage changes in the cable power supply of four different transoceanic cables during time periods of high solar activity. We find a strong correlation between the strength of the high-frequency geomagnetic field at the landing sites of the systems and the line voltage change. We also uncover that these two quantities exhibit a near-linear power law scaling behavior that allows us to estimate the effects of once-in-a-century CME events. Our findings reveal that long-haul submarine cables, regardless of their length and orientation, will not be damaged during a solar superstorm, even one as large as the 1859 Carrington event.

## Introduction

Coronal Mass Ejections (CMEs), commonly known as solar storms, are a space weather phenomena that involves the directional emission of solar matter into interplanetary space (Forbes, 2000). If the Earth happens to be in the way of one of these emissions, the interaction of the Sun's electrically charged particles with the Earth's magnetosphere will lead to a rapid disturbance on the geomagnetic field that will, in turn, generate a geoelectric field. This geoelectric field propagates through the conducting interior of the Earth and produces voltages across grounded transmission lines that lead to Geomagnetically Induced Currents (GICs; Pirjola, 2007; 2009). Depending on the size of the solar storm, the associated GIC can cause operational interference and damage to power transmission networks that result in extensive outages (Knipp et al., 2016; Love et al., 2016).

Similarly to earthquakes, solar storms are difficult to predict and small-magnitude events occur constantly in time. The largest solar storm ever documented, the 1859 Carrington event, caused severe damage to the telegraph network (Carrington, 1859;

Hodgson, 1859). More than a century later, a large solar storm in 1989 produced a GIC strong enough to cause the collapse of the Hydro-Quebec power grid in Canada (Bolduc, 2002). In 2012, an even more powerful CME tore through the Earth's orbit but missed the Earth by one week. It is hypothesized that if that solar emission had hit Earth's atmosphere, the resulting geomagnetic storm would have registered a perturbation comparable to the Carrington event, and at least twice as strong as the 1989 Quebec blackout (Baker et al., 2013). An important point to note is that all these events occurred during time periods of high solar activity, which is observed to trace an 11-year peak cycle (Hathaway, 2015; Figure S1). Currently, the Sun is emerging from the quiescent period of this cycle, suggesting that the next few years will see significantly increased risks from CMEs. Recent estimates place the probability of experiencing another Carrington-like event within the next decade at 2-12% (Riley, 2012; Moriña et al., 2019).

The implications of space weather events on space and terrestrial infrastructures have been actively studied over the past decades. Effects on power grids (Abda et al., 2020, Kappenman, 2001), railway networks (Belov et al., 2007), aviation (Jones et al., 2005), satellite communications (Green et al., 2017), and global navigation satellite systems (Sreeja, 2016) have been explored by researchers and policy makers to minimize economical and societal impacts of black-swan solar events (NRC, 2008, Cannon et al., 2013, Eastwood et al., 2017). Analyses on the impact of geomagnetic storms in transoceanic telecommunication links have also been conducted since the late 50s following disruptive events on telegraph cables (Axe, 1968; Anderson et al., 1974). Medford et al. (1989), for example, documented the effects of the 1989 Quebec blackout event on TAT-8, the first fiber optic transatlantic cable, and reported rapid line voltage variations during the intense portion of the geomagnetic disturbance. Other similar studies focused on investigating the magnitude of the large-scale potential differences that can be induced within the Earth by fluctuations of the geomagnetic field, and analyzed their impact on different submarine interconnected systems (e.g. Lanzerotti et al., 1992, Lanzerotti et al., 1995, Lanzerotti, 2001). In 2019, Enright et al., collected line voltage measurements from twelve legacy (installed before 2015) and three modern submarine cable systems during active geomagnetic intervals, and provided updated design recommendations to mitigate the risk of power fluctuations in undersea cable networks.

Most recently, Jyothi (2021) presented an analysis highlighting the risk of partitioning of the Internet in case of a once-in-a-century CME event. In more detail, it is pointed out that large GICs could surge through subsea telecommunication cables, which are designed to operate at a constant 1 A current, and burn out the repeaters that boost the optical signals. Submarine cables take 2-3 years to construct and, therefore, the destruction of multiple cables would have a severe and long-lasting impact.

Here, we consider the various geophysical factors of ocean environments, and the design of modern undersea cable networks, to assess the vulnerability of the submarine Internet backbone infrastructure to GICs. We start by examining continuous voltage logs from the Power Feeding Equipment (PFE) of four separate transoceanic cable systems and comparing them with synchronous readings from magnetometers operating near the landing sites of the cables. We then quantify the strength of space-related perturbations in the magnetometer data and correlate it with the voltage flowing through the subsea systems at those points in time. Lastly, we examine the relationship between these two quantities and extrapolate it to define a threshold for the operational capabilities of submarine cable networking. We conclude that design and operating conventions of modern subsea systems protect them from large solar storms, including those similar to the Carrington event.

## Engineering Considerations for Submarine Cable Systems

Long-haul submarine systems are powered from PFEs housed in their cable landing stations, one at each end. While the fiber strands that carry the digital data do not use any electrical power, signal attenuation along the fiber links requires the optical signals to be boosted at regular intervals via optical amplifiers (i.e. repeaters) that do consume power. The PFEs at each landing station feed the required DC current, which is then transmitted via a copper conductor embedded within the cable (Figure 1A-B). This conductor makes subsea repeaters vulnerable to damage from GICs.

Because submarine systems are designed to operate continuously for 25 years, high availability and resilience governs every design aspect of the subsystems. The repeaters' power regulation circuitry is designed to withstand current surges by deploying a series of cascaded Zener diodes, typically rated at ~700 amp, to shunt any excessive current. This power regulation design protects the amplifier electronics from line current perturbations and protects them from a reverse polarity situation (Takehira, 2016). The powering of the submarine system is strictly informed by the requirement that the PFE maintain a fixed constant current to the line. As a result, the PFEs are continuously adjusting the line voltage in response to environmental and physical changes to the submarine system to maintain that constant current (Figure 1C). Monitoring the overall system voltage thus provides a means to observe changes in the properties of the submarine line and/or the environment in which the network is operating.

In terms of headroom, PFE systems include an Earth Potential Allowance (EPA) variance, which is a safety factor used when calculating the maximum system voltage (e.g. Enright, 2019). Depending upon the operational resilience that is desired for a particular system or network, the EPA normally ranges from 0.05 V/km to 0.1 V/km. This EPA is added to the system design voltage – the voltage necessary to power the repeaters and the voltage drop across the resistive cable – to provide margin in the event changes occur to the environment in which the whole wet plant operates. The PFE at any one end of the system is then dimensioned to have sufficient voltage headroom to power the cable and repeaters but also to have this EPA safety margin. A typical trans-Atlantic system, for instance, has an EPA of 650-700 V, and this extra capacity contributes to the overall max system design voltage. The majority of modern Atlantic submarine systems are characterized by a total  system voltage of ~11,000 V and are thus equipped with PFEs that can deliver almost 12,000 V at each end of the system to allow the system to be powered from one end in the event of PFE fault.

Submarine systems also include another resilience design feature that supports continued availability in the unusual event of PFE failure. The industry standard powering design convention for submarine systems is to operate in Double End Feeding (DEF) configuration, with each PFE providing half of its maximum voltage so that the two PFE share the load simultaneously and the PFE are only supplying half of their powering capability. In the event of a single PFE failure, the inherent capacitance of the +6,000-km line (for the typical Atlantic system) provides a collapsing interval before system power falls and line current drops, allowing the remaining operational PFE to ramp up to twice its normal DEF voltage level to maintain the designed constant line current value. The continuous adjustment of voltage to maintain a constant current, combined with the large allowable voltage range, protects the submerged plant from unexpected environmental events such as large solar storms. In DEF mode they have essentially 50% power headroom plus EPA variance to add to the total system headroom.

## Data and Analysis

In principle, the system voltage is determined by the voltage drop across the conductor, the voltage drop across each repeater, and the potential difference between the landing sites (Figure 1C). Assuming that the physical properties of a system are constant through time, we can, therefore, attribute any re-adjustments in the voltage of the system to changes in the Earth's potential (Medford et al., 1989; Lanzerotti et al., 1995). This potential difference is determined by the combined effects of environmentally-induced magnetic field fluctuations and the conductive properties of the ground at the opposite ends of the fiber (i.e., coastal effects). As an example, Figure 2 shows the broadband 5-day voltage log of a PFE that supplies power to a 10,000 km long submarine cable connecting Santos, Brazil and Florida, United States. While the overall readings from the PFE are centered around the nominal voltage of the system (~6,872 V), the measurements show a 12-hour periodic behavior with a 4 V amplitude. The magnetometer data recorded at the VSS observatory in Vassouras, Brazil, approximately 300 km northeast of the cable's PFE, reveals the same periodic pattern emerges in the E-W components of the geomagnetic field. Both of these quantities are in close phase with the semi-diurnal tidal cycle at the Brazilian shore as a result of the conductive sea water pulling and pushing the local magnetic field lines along with the tides, introducing fluctuations on the local geomagnetic field and, in the process, inducing multiple GICs that flow through the cable powering path (Hewson-Browne, 1973; Petereit et al., 2019).

During the same time period that is shown in Figure 1, a small geomagnetic storm occurred on September 8, 2021. According to the NOAA Space Weather Prediction Center (SWPC), this storm had a geomagnetic K-index of 4, which is expected to cause mild fluctuations on power grid voltages. The signature of this storm is captured by both the VSS magnetometer and the system's PFE, producing approximately a 5 V increase on the latter. Considering the length of the cable, this event produced a potential difference of 0.0005 V/km, which is well-below the typical 0.1 V/km EPA. In this study, we focus on PFE voltage changes that are associated with this type of GICs.

We compiled one month's worth of PFE voltage logs of two trans-Atlantic and two trans-Pacific telecommunication cables (Figure 3). In light of the double feed configuration, we add the voltage values from both PFEs to assess the overall systems' response to GICs. The date range of the datasets and the characteristics of the systems are presented in Table S1. For the magnetometer data, we use recordings of the horizontal components (E-W and N-S) of the Earth's magnetic field from the INTERMAGNET observatories that are closest to each of the cable's landing sites (Kerridge, 2001). Note that we do not consider the vertical component of the magnetic field since we assume a simple model of GIC generation, where the current is driven by the horizontal geoelectric field that is induced by a plane wave propagating vertically downwards (e.g. Pirjola 2000, 2002).

Because the intensity of a GIC depends on the rate of change of the geomagnetic field, rather than its absolute amplitude, we focus on the mHz frequency range of the magnetometer data, high-pass filtering the signals at 0.001 Hz (Text S1). This processing step is also designed to remove any tidal signature in the timeseries while retaining the frequencies of the geomagnetic field that have been reported to generate hazardous GICs (Barnes, 1991; N.A.E.R.C, 2014). We then average the amplitude spectrum of the horizontal geomagnetic fields, and use its frequency-integrated representation, henceforth referred to as Jo-unit, as a metric of the magnetic perturbation that generates a GIC in a submarine cable powering circuit. More succinctly, we express the Jo-unit metric as

$$ Jo = \int_{f1}^{f2} \frac{1}{2} (ASD_{E-W} + ASD_{N-S}) df, $$

where ASD represents the amplitude spectral density of the E-W and N-S geomagnetic fields, and (f1, f2) the frequency range of integration, which, for this case, goes from 0.001-Hz to 0.008-Hz as limited by the Nyquist sampling rate. For the PFE measurements, we only subtract their nominal voltage values, apply a 6-hour highpass filter to remove the tidal component, and estimate their Hilbert envelope.

Figure 4 shows a 10-day window of the PFE voltage measurements for a 10,500 km long cable connecting Los Angeles, California, and Valparaiso, Chile, together with the Jo-unit values from the magnetometers that are near the PFEs. During this time a class-M solar flare occurred and reached Earth's atmosphere, producing a geomagnetic storm with a K-index of 7. Both the amplitude of the local high-frequency magnetic fields and the voltage that is flowing through the cable powering path show a notable temporal agreement, with the geomagnetic storm producing a maximum voltage increase of 10 V on both PFEs. We attribute the similarity of the response of the two PFEs to the fact that, while spatially dependent, geomagnetic storms are global events that manifest as an almost synchronous change in the geomagnetic field at both ends of the sysyem (e.g. Wert et al, 2022). To obtain a general measure of the system response to space-related GICs, we average the PFE voltages and the Jo-unit amplitudes that are measured at the two ends of the system.

Figure 5 shows the line voltage increase as a function of Jo-units for all events observed during the measurement time period. To expand our dataset further, we include the measurement of the March 1989 geomagnetic storm as sensed by the TAT-8 trans-Atlantic cable at the North America end (maroon star in Figure 5; Medford, 1989). This event is one of the largest CMEs ever recorded by any ground-based instrument, producing a peak voltage that is ~300 V above the nominal value of the system. The associated Jo-unit amplitude for this measurement is extracted from the FRD magnetometer in Fredericksburg, USA, ~300 km away from the location of the PFE (Figure S3).

## Results

Regardless of the length and orientation of the cable, we find that the line voltage variability shows a self-similar power law scaling behavior that can be described by the functional form:

$$V^+ = a \cdot Jo^b,$$

where $V^+$ is the average voltage increase of the two PFEs, and $Jo$ is the average Jo-unit amplitude at the landing sites of the cable. For the data points in Figure 5, the best-fitting values of Eq. 2 are given by $a$=0.44±0.0197 and $b$=0.97±0.0071, with a coefficient of determination, $R^2$, of 0.97. This result suggests that there is a close-to linear relationship between the strength of a GIC on a subsea cable and the high-frequency fluctuations of the magnetic field at its grounding ends.

To validate our model, we exclude the measurement of the March 1989 geomagnetic storm and then perform the same power law regression on 500 bootstrapped families of the remaining data points (Figure S4). We find that the TAT-8 measurement falls within the 2σ range of solutions, suggesting that our model is robust enough to have approximated the average effects of the March 1989 geomagnetic storm on a submarine network. The deviations in the measurements, and variance in the fits, can be explained by a wide variety of factors including the intrinsic variations in the impedance properties of the Earth's crust and mantle, the differences in distance between the magnetometers and the PFEs, and the limited scale of the ionosphere spatial variations during the geomagnetic storms. Because we are deriving a potential difference proxy directly from magnetometer data, our framework does not need to consider any latitude-dependence of the system.

While it is generally expected that the voltage change of a terrestrial system during a fluctuation of the geomagnetic field is proportional to the length of the conductor, we find no clear correlation between these two quantities for submarine cables. This observation is consistent with Enright (2019), who documented no relation between voltage change and subsea cable length during periods of high solar activity. We attribute this dissociation to the high-resistance of the cable and to the presence of the conductive ocean layer, which strongly attenuates the amplitude of the high-frequency magnetic pulsations from the incident CMEs and, consequently, prevents them from inducing large geoelectric fields along the deep water sections of cables (Text S2). As a result, the major contribution to the voltage change in a submarine system originates from the geoelectric fields that are induced at the coast only and not anywhere along the fiber route itself. From this finding, we suggest that the correct metric for establishing a conservative submarine EPA should be Volts and not V/km.

## Discussion

With a general relation describing the voltage increase of a submarine system as a function of the strength of a geomagnetic perturbation, we can now estimate the effects of a solar superstorm. The Carrington event, for instance, is hypothesized to have produced magnetic perturbations approximately 2-to-3 times higher than the March 1989 geomagnetic storm. Assuming similar ground impedances as the ones in the landing sites of the fibers analyzed in this study, a magnetic perturbation of such magnitude would generate a voltage increase in a subsea system that is just below 1,000 V. Taking the Atlantic E-W #1 system as an example, each of its PFEs operate with a 5,226 V nominal voltage and are designed to hold an individual maximum load of 12,000 V, providing a headroom of approximately 6,000V. The cable is 6,600-km long and the average resistance of its power line is of 1.3 Ω/km. These values suggest that a Carrington-like event would introduce a 0.11 A current change on the system. While the magnitude of this GIC does exceed the standard EPA, the head-room allocated through the double-end power feed design is still sufficient to compensate for this extra current. In agreement with Medford, (1989) our results indicate that the various subsea system design features, including the PFE dimensioning, are conservative enough to protect submarine telecommunication cables from the effects of large and super solar storms events.

## Conclusion

We measured voltage changes in the cable power supply of four different transoceanic cables to find a strong correlation ($R^2 = 0.97$) between the strength of the high-frequency component of the geomagnetic field and the line voltage change. We also uncover a near-linear power law scaling behavior that allows estimating the effects of once-in-a-century CME events. Our findings reveal that long-haul submarine cables, regardless of their length and orientation, will not be damaged during a solar superstorm, even one larger than the 1859 Carrington event. In fact, submarine cable infrastructure will weather a 10x stronger storm, in which case PFE voltage will approach the limit, and gently shut down to be manually turned on after the storm.

We explain the ability of cable systems to sustain operation by their dual power feed design. Voltage controllers, which are designed to provide stabilized current under extreme conditions, can compensate for geomagnetically induced currents. Submarine cables are further protected from GICs since the majority of the cables' path is at substantial ocean depths. The low resistivity of sea-water acts as a strong attenuator of electromagnetic waves, protecting deeply submerged segments from GICs. As a result, the magnitude of the induced currents does not increase with cable length so that even very long submarine cables will not be endangered by large CME events. Thus, we conclude that the primary risk to internet infrastructure resulting from large CME events lies in their impact on terrestrial power grids.

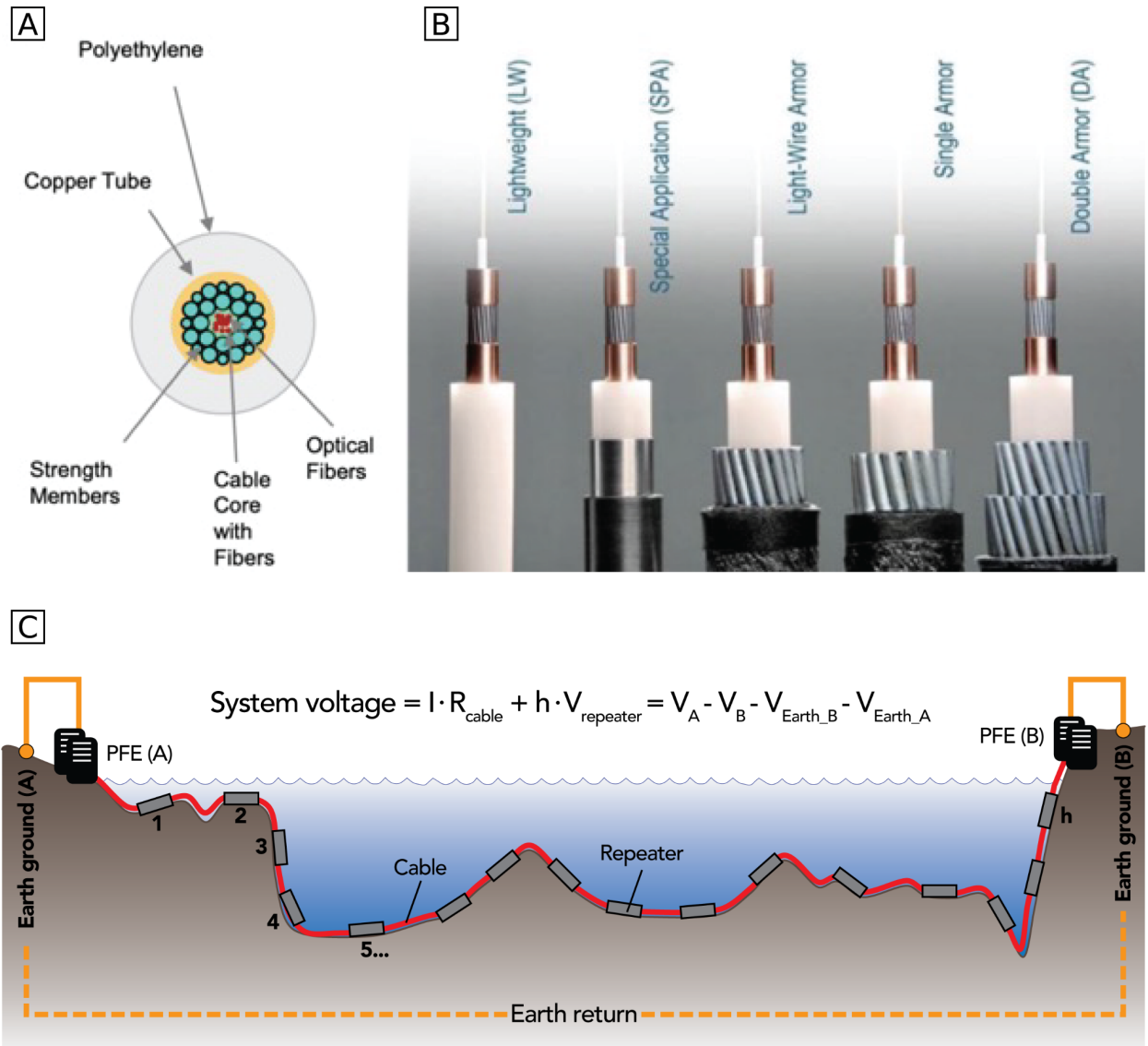

**Figure 1.** Structure and types of the telecommunication fiber cables with copper conductor power feed from two shores. (A) Lightweight (LW) optical fiber cable with steel strand wire configuration. The interlocking strength members resist external pressure up to 8,000 m water depth. LW cable (typically more than 90% of cable length is LW) has external diameter 17 mm and weight in air 4.8 kN/km. The cable core with fibers consists of a tube filled with gel. (B) Armored part of the cable (typically 5-10%) consists of Special Application, Light-Weight Armor, and Double Armor cable. (C) Electric circuit to power feed subsea repeaters 1, 2, 3, 4, 5...h includes ohmic loss I*$R_{cable}$, two redundant power feed equipment sources of stabilized DC current (PFE(A), PFE(B)), and Earth return with Earth ground(A), Earth ground(B). Panels (A) and (B) are taken from Chesnoy (2015).

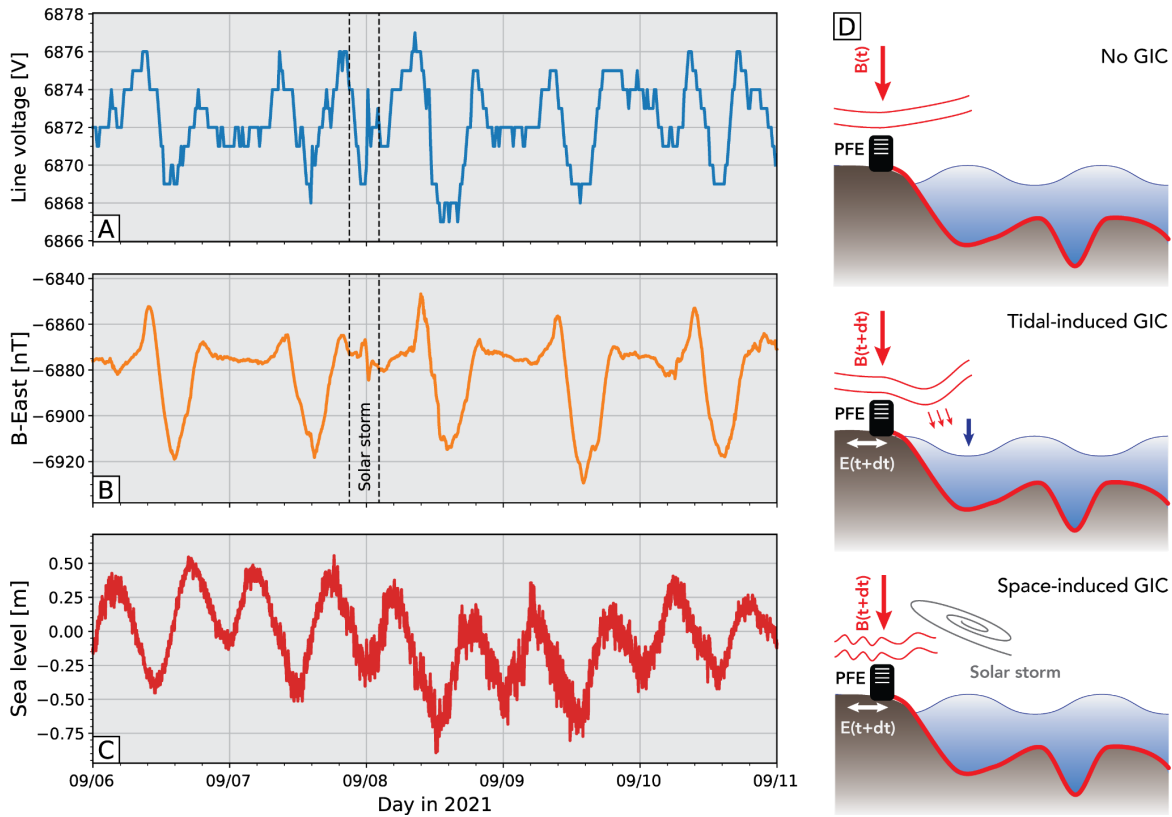

**Figure 2.** Comparison between a 5-day broadband record of A) line voltage of a 10,000 km long submarine fiber connecting Santos, Brazil and Florida, United States, B) E-W component of the geomagnetic field at the VSS observatory in Vassouras, Brazil, 300 km northeast of the cable's PFE, and C) Relative sea level measured by the Imbituba tide gauge in Brazil, 500 km south of the cable's PFE. All three time series exhibit the dominant 12-periodicity that is characteristic of the semidiurnal tidal cycle. The dashed black lines in (A) and (B) mark the time of occurrence of a small geomagnetic storm on September 8, 2021. Figure S2 shows the location of the PFE, the magnetometer, and the tide gauge. D) Shows a schematic diagram of the mechanisms responsible for generating the line voltage fluctuations in (A). For a given time, t, the local geomagnetic field, B, is constant, no geoelectric field, E, is generated, and no current is induced on the fiber (top). The conductive sea water interacts with the local geomagnetic field and produces multiple long-period GICs (middle). A solar storm generates a high-intensity flux on the geomagnetic field and produces a relatively short duration GIC (bottom).

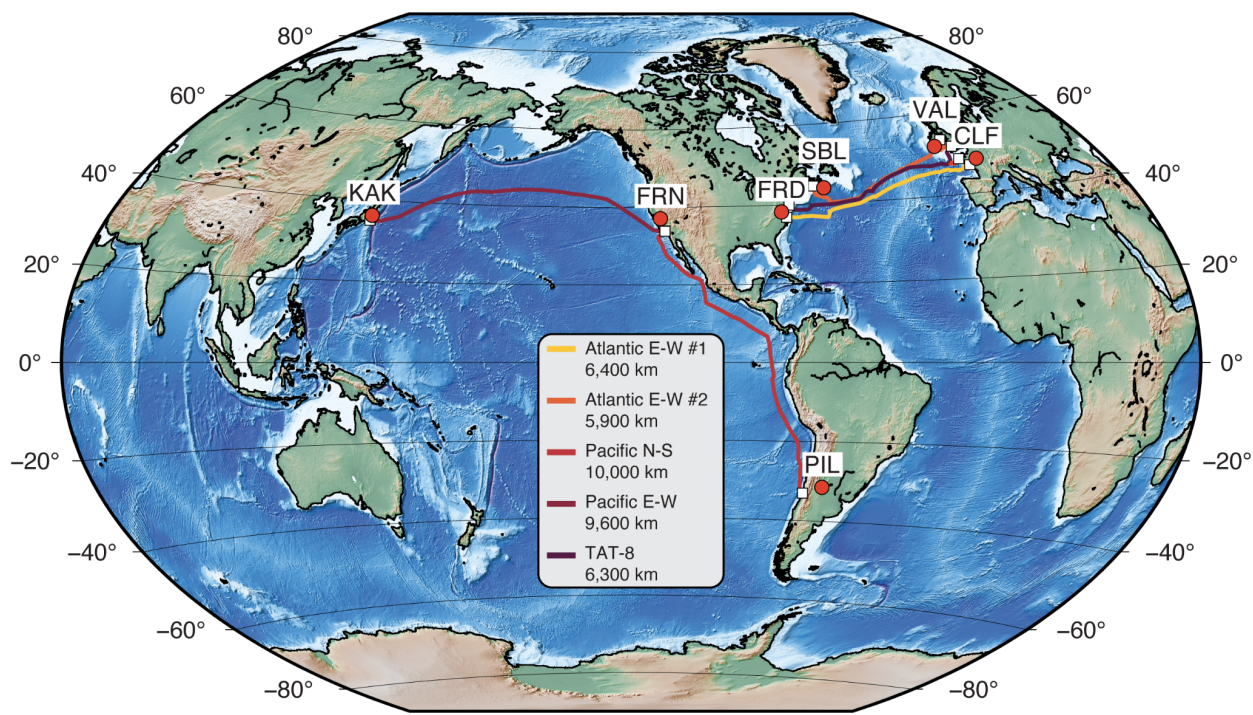

**Figure 3.** Geographical location of the transoceanic links and ground-based magnetometers (red circles) used in this investigation. The name of each magnetic observatory is specified on top of each marker. The length of each system is specified in the legend.

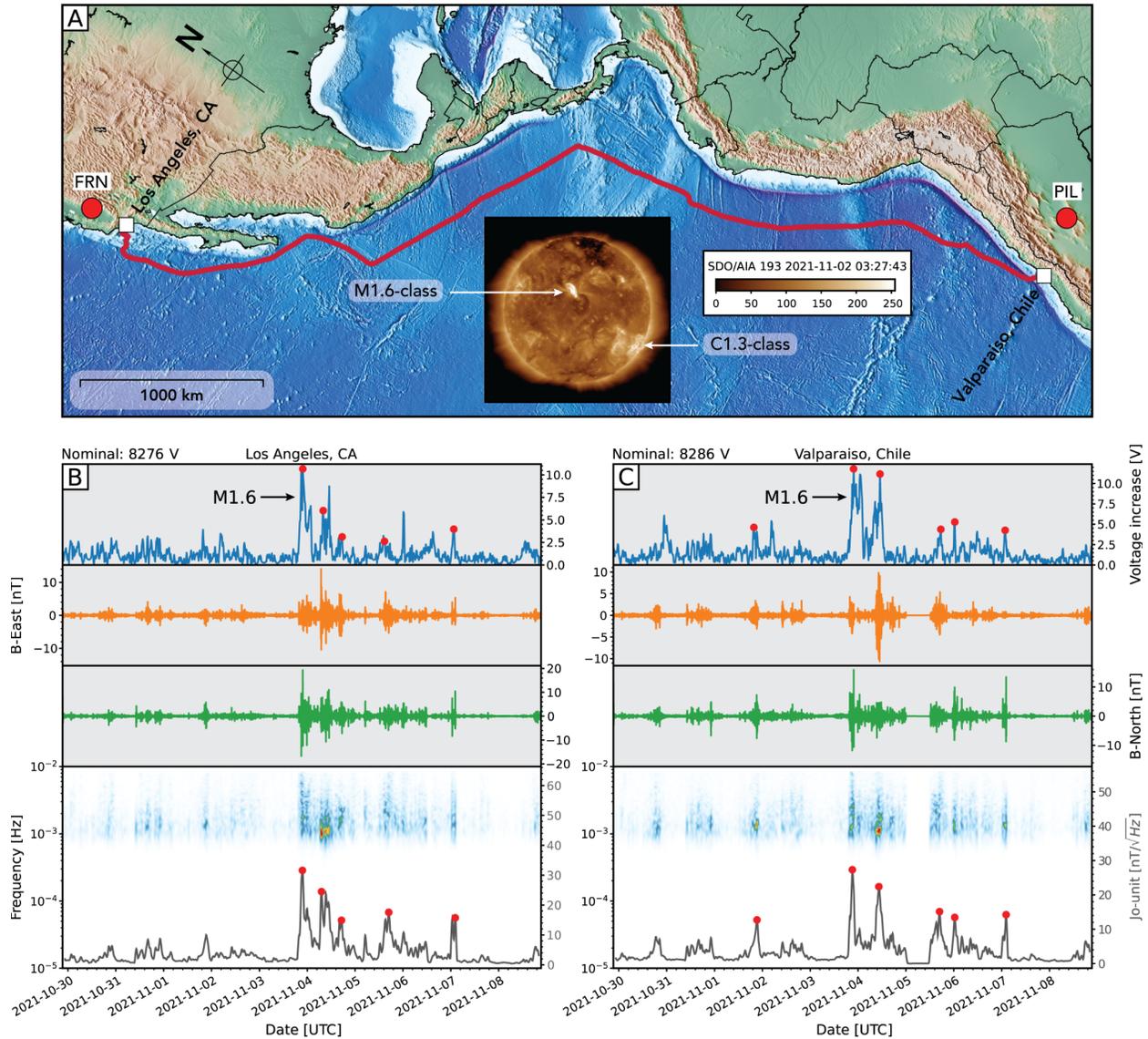

**Figure 4.** PFE voltage increase and magnetometer measurements for the trans-Pacific N-S system during a geomagnetic storm. A) Shows the geographical location of the system and the location of the two magnetometers. The inset figure shows the coronal image of the Solar flare recorded by the SDO/AIA 193 channel on 2 November of 2021. Two flares can be identified at that time: (1) a C1.3 class located in the bottom right region, and (2) a M1.6 class located at the center of the image. B) Shows the line voltage measurements as logged by Los Angeles PFE (blue timeseries), the east and north components of the geomagnetic fields (yellow and green timeseries, respectively) and their Jo-unit representation (gray timeseries), as recorded by the FRN magnetometer. The average amplitude spectra of the magnetic signals are also shown in the bottommost panel. The red circles indicate the times where the magnetometer data shows peaks in their Jo-unit representation. C) Shows the same

information as in (B), but for the Chile PFE and the PIL magnetometer. It is worth noting in (B) and (C) that both PFEs are sensitive enough to even capture the multiple minor radiation storms that typically follow a large magnitude solar event.

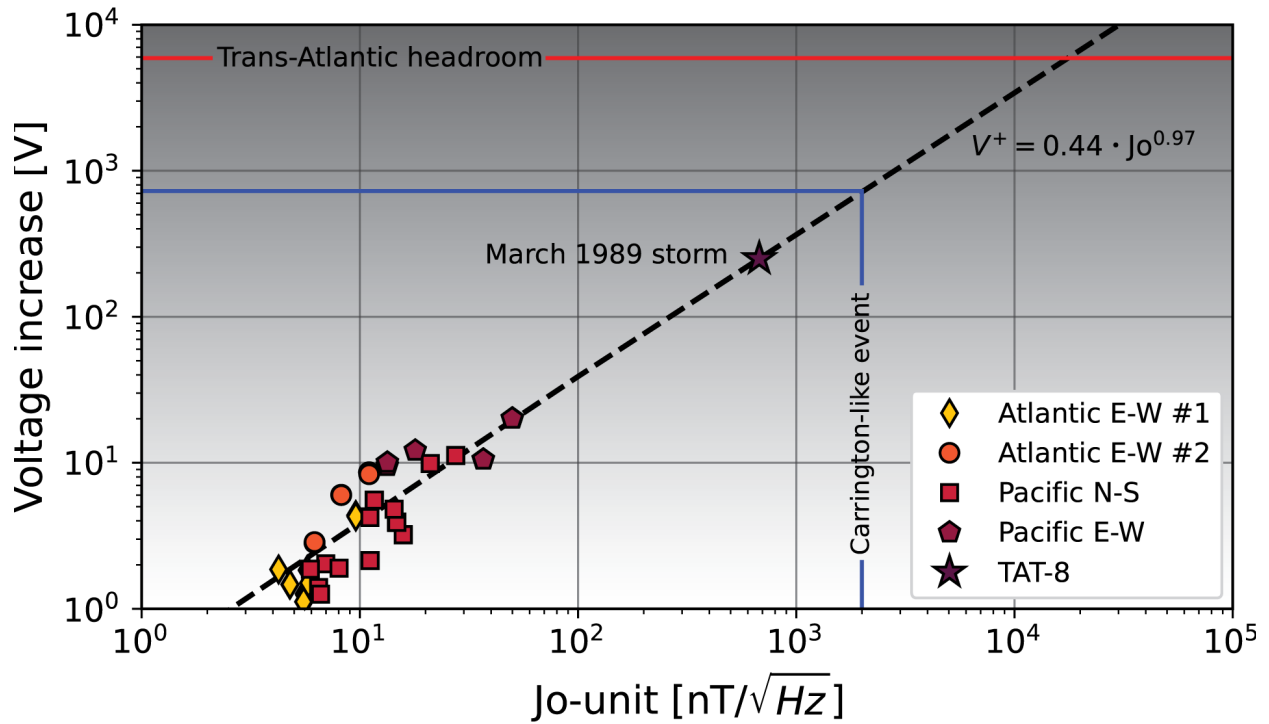

**Figure 5.** PFE voltage increase as a function of the Jo-unit amplitudes at the landing site of the systems for all events observed during the measurement period. The maroon star marks measurement for the March 1989 storm (Medford, 1989). The black dashed line marks the best-fitting result of the regression. The blue line marks the expected Jo-unit amplitude of a Carrington-like event and its associated line voltage increase. The red line marks the voltage headroom for a typical trans-Atlantic system.

# References

1. Forbes, T. G. (2000). A review on the genesis of coronal mass ejections. *Journal of Geophysical Research: Space Physics*, *105*(A10), 23153-23165.

2. Pirjola, R. (2007). Calculation of geomagnetically induced currents (GIC) in a high-voltage electric power transmission system and estimation of effects of overhead shield wires on GIC modeling. *Journal of atmospheric and solar-terrestrial physics*, *69*(12), 1305-1311.

3. Pirjola, R. (2009). Properties of matrices included in the calculation of geomagnetically induced currents (GICs) in power systems and introduction of a test model for GIC computation algorithms. *Earth, planets and space*, *61*(2), 263-272.

4. Knipp, D. J., Ramsay, A. C., Beard, E. D., Boright, A. L., Cade, W. B., Hewins, I. M., ... & Smart, D. F. (2016). The May 1967 great storm and radio disruption event: Extreme space weather and extraordinary responses. *Space Weather*, *14*(9), 614-633.

5. Love, J. J., Coïsson, P., & Pulkkinen, A. (2016). Global statistical maps of extreme-event magnetic observatory 1 min first differences in horizontal intensity. *Geophysical Research Letters*, *43*(9), 4126-4135.

6. Carrington, R. C. (1859). Description of a singular appearance seen in the Sun on September 1, 1859. *Monthly Notices of the Royal Astronomical Society*, *20*, 13-15.

7. Hodgson, R. (1859). On a curious appearance seen in the Sun. *Monthly Notices of the Royal Astronomical Society*, *20*, 15-16.

8. Bolduc, L. (2002). GIC observations and studies in the Hydro-Québec power system. *Journal of Atmospheric and Solar-Terrestrial Physics*, *64*(16), 1793-1802.

9. Baker, D. N., Li, X., Pulkkinen, A., Ngwira, C. M., Mays, M. L., Galvin, A. B., & Simunac, K. D. C. (2013). A major solar eruptive event in July 2012: Defining extreme space weather scenarios. *Space Weather*, *11*(10), 585-591.

10. Hathaway, D. H. (2015). The solar cycle. *Living reviews in solar physics*, *12*(1), 1-87.

11. Riley, P. (2012). On the probability of occurrence of extreme space weather events. *Space Weather*, *10*(2).

12. Moriña, D., Serra, I., Puig, P., & Corral, Á. (2019). Probability estimation of a Carrington-like geomagnetic storm. *Scientific reports*, *9*(1), 1-9.

13. Abda, Z. M. K., Ab Aziz, N. F., Ab Kadir, M. Z. A., & Rhazali, Z. A. (2020). A review of geomagnetically induced current effects on electrical power systems: Principles and theory. IEEE Access, 8, 200237-200258.

14. Kappenman, J. G. (2001). An introduction to power grid impacts and vulnerabilities from space weather. *Space Storms and Space Weather Hazards*, 335-361.

15. Belov, A. V., Gaidash, S. P., Eroshenko, E. A., Lobkov, V. L., Pirjola, R., & Trichtchenko, L. (2007, June). Effects of strong geomagnetic storms on Northern railways in Russia. In *2007 7th International Symposium on Electromagnetic Compatibility and Electromagnetic Ecology* (pp. 280-282). IEEE.

16. Jones, J. B. L., Bentley, R. D., Hunter, R., Iles, R. H. A., Taylor, G. C., & Thomas, D. J. (2005). Space weather and commercial airlines. Advances in Space Research, 36(12), 2258-2267.

17. Green, J. C., Likar, J., & Shprits, Y. (2017). Impact of space weather on the satellite industry. Space Weather, 15(6), 804-818.

18. Sreeja, V. (2016). Impact and mitigation of space weather effects on GNSS receiver performance. Geoscience letters, 3(1), 1-13.

19. National Research Council. (2008). Severe space weather events: Understanding societal and economic impacts: A workshop report.

20. Cannon, P., Angling, M., Barclay, L., Curry, C., Dyer, C., Edwards, R., ... & Underwood, C. (2013). Extreme space weather: impacts on engineered systems and infrastructure. Royal Academy of Engineering.

21. Eastwood, J. P., Biffis, E., Hapgood, M. A., Green, L., Bisi, M. M., Bentley, R. D., … & Burnett, C. (2017). The economic impact of space weather: Where do we stand?. Risk Analysis, 37(2), 206-218

22. Axe, GA (1968. The effects of the earth's magnetism on submarine, Post Office Electrical Engineers' Journal, 61(1), 37-43

23. Anderson III, C. W., Lanzerotti, L. J., & Maclennan, C. G. (1974). Outage of the L4 system and the geomagnetic disturbances of 4 August 1972. Bell System Technical Journal, 53(9), 1817-1837.

24. Medford, L. V., Lanzerotti, L. J., Kraus, J. S., & Maclennan, C. G. (1989). Transatlantic earth potential variations during the March 1989 magnetic storms. *Geophysical Research Letters*, *16*(10), 1145-1148.

25. Lanzerotti, L. J., Sayres, C. H., Medford, L. V., Kraus, J. S., & Maclennan, C. G. (1992). Earth potential over 4000 km between Hawaii and California. Geophysical research letters, 19(11), 1177-1180.

26. Lanzerotti, L. J., Medford, L. V., Maclennan, C. G., & Thomson, D. J. (1995). Studies of large-scale Earth potentials across oceanic distances. *AT&T technical journal*, *74*(3), 73-84.

27. Lanzerotti, L. J. (2001). Space weather effects on technologies. *Washington DC American Geophysical Union Geophysical Monograph Series*, *125*, 11-22.

28. Enright, M. (2019, April). Solar Cycles and Voltages in Earth Potential. *SubOptic2019 Conference*.

29. Jyothi, S. A. (2021). Solar superstorms: planning for an internet apocalypse. In *Proceedings of the 2021 ACM SIGCOMM 2021 Conference* (pp. 692-704).

30. Chesnoy, J. (2015). Undersea fiber communication systems. Academic press.

31. Takehira, K. (2016). Submarine system powering. In *Undersea Fiber Communication Systems* (pp. 381-402). Academic Press.

32. Hewson-Browne, R. C. (1973). Magnetic effects of sea tides. *Physics of the Earth and Planetary Interiors*, *7*(2), 167-186.

33. Petereit, J., Saynisch-Wagner, J., Irrgang, C., & Thomas, M. (2019). Analysis of Ocean Tide-Induced Magnetic Fields Derived From Oceanic In Situ Observations: Climate Trends and the Remarkable Sensitivity of Shelf Regions. *Journal of Geophysical Research: Oceans*, *124*(11), 8257-8270.

34. Kerridge, D. (2001, December). INTERMAGNET: Worldwide near-real-time geomagnetic observatory data. In *Proceedings of the workshop on space weather, ESTEC* (Vol. 34).

35. Pirjola, R. (2000). Geomagnetically induced currents during magnetic storms. *IEEE transactions on plasma science*, *28*(6), 1867-1873.

36. Pirjola, R. (2002). Review on the calculation of surface electric and magnetic fields and of geomagnetically induced currents in ground-based technological systems. *Surveys in geophysics*, *23*(1), 71-90.

37. Barnes, P. R., Rizy, D. T., Mcconnell, B. W., Taylor Jr, E. R., & Tesche, F. M. (1991). *Electric utility industry experience with geomagnetic disturbances* (No. ORNL-6665). Oak Ridge National Lab., TN (United States).

38. North American Electric Reliability Corporation. (2014). Benchmark Geomagnetic Disturbance Event Description.

39. Wert, J. L., Dehghanian, P., Zhang, A., Stevens, M., Guthrie, R., Snodgrass, J., Shetye K. S., Overbye T. J., Davis K. R. & Gannon, J. (2022). Analysis of Magnetometer Data from a Strong G3 Geomagnetic Disturbance.

## Supplementary material

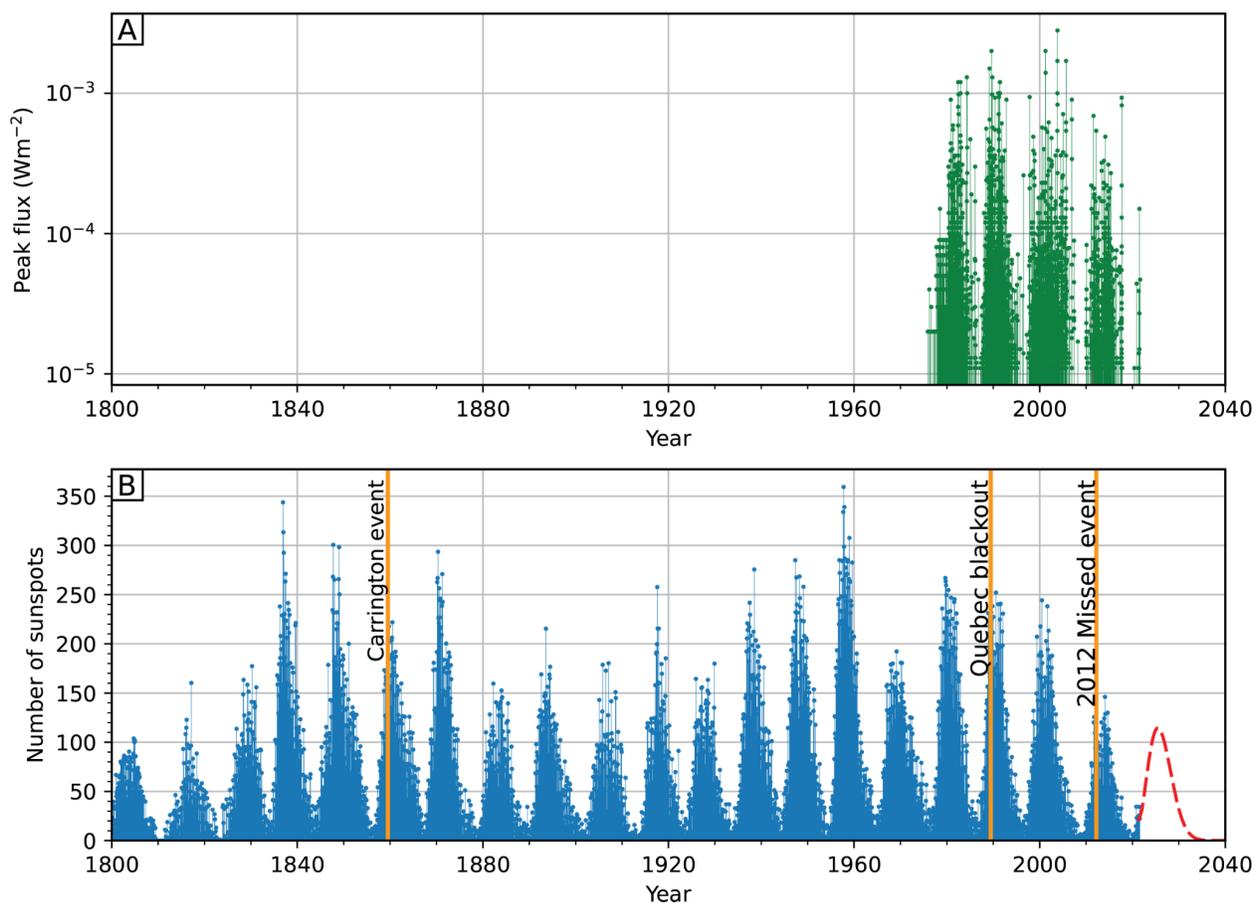

**Figure S1**. Flux levels of solar flare radiation recorded by the X-ray sensors on board of the GOES series of satellites (A) and Solar cycle in terms of number of sunspots (B). The red dashed line marks the prediction for how the sunspot number will evolve in the next 20 years. The yellow lines mark the timing of the 1859 Carrington event, the 1989 Quebec blackout event, and the 2012 event. Data source: U.S. Dept. of Commerce, NOAA, Space Weather Prediction Center (SWPC).

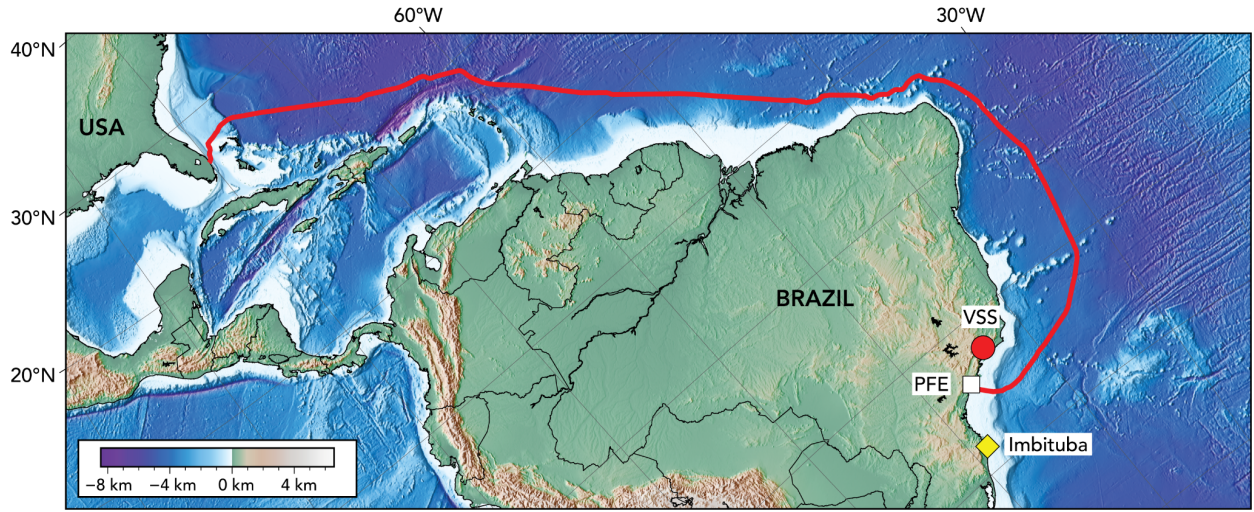

**Figure S2**. Geographical location of the transoceanic link (red line), the ground-based magnetometer (VSS; red circle), and the tide gauge (Imbituba; yellow diamond), whose data is shown in Figure 1.

**Text S1. Details on the Jo-unit scale**

It is generally well-known that the intensity of a geoelectric field mostly depends on the conductive properties of the Earth's crust and mantle, and the rate-of-change of the geomagnetic field. The process of estimating the impact of a solar storm on power transmission networks thus requires detailed knowledge of the local geologic structure and the strength of the geomagnetic fluctuations along and across each of the fiber lines. For subsea systems, however, the complexity of this problem is significantly reduced due to the presence of the ocean layer above them, which largely attenuates the electromagnetic fields as they travel down from the surface (Text S2). This property allows us to neglect the space-induced electromagnetic effects in the deep-ocean sections of the fiber, and associate, to first-order, the strength of a GIC to the geoelectric fields that are induced in the vicinity of the landing sites of the systems only.

Without any available electric impedance values along the shorelines, or a more developed understanding of the non-linear coastal effects on geoelectric fields (e.g. Wang et al., 2022), it is only possible to derive an empirical metric that maps the size of a geomagnetic perturbation to the amplitude of a GIC in a submarine system. For this purpose, we derive the Jo-unit scale, which is a measure of the strength of the geomagnetic field fluctuations at the landing sites of a submarine system. To obtain this value we apply a fourth-order Butterworth high-pass filter with a 0.001 Hz cutoff to the horizontal components of the geomagnetic field, average their amplitude spectrum, and integrate it across the frequency dimension (Figure S3). The cutoff frequency of the high-pass filter was determined based on previous investigations that associate hazardous GIC intensities to geomagnetic field variations above that frequency. An interesting thing to note is that, by applying this high-pass filter to the geomagnetic data, we arrive at a time-differentiated version of the signals (i.e. taking the temporal derivative of a signal increases the amplitude of the high frequencies relative to the low frequencies).

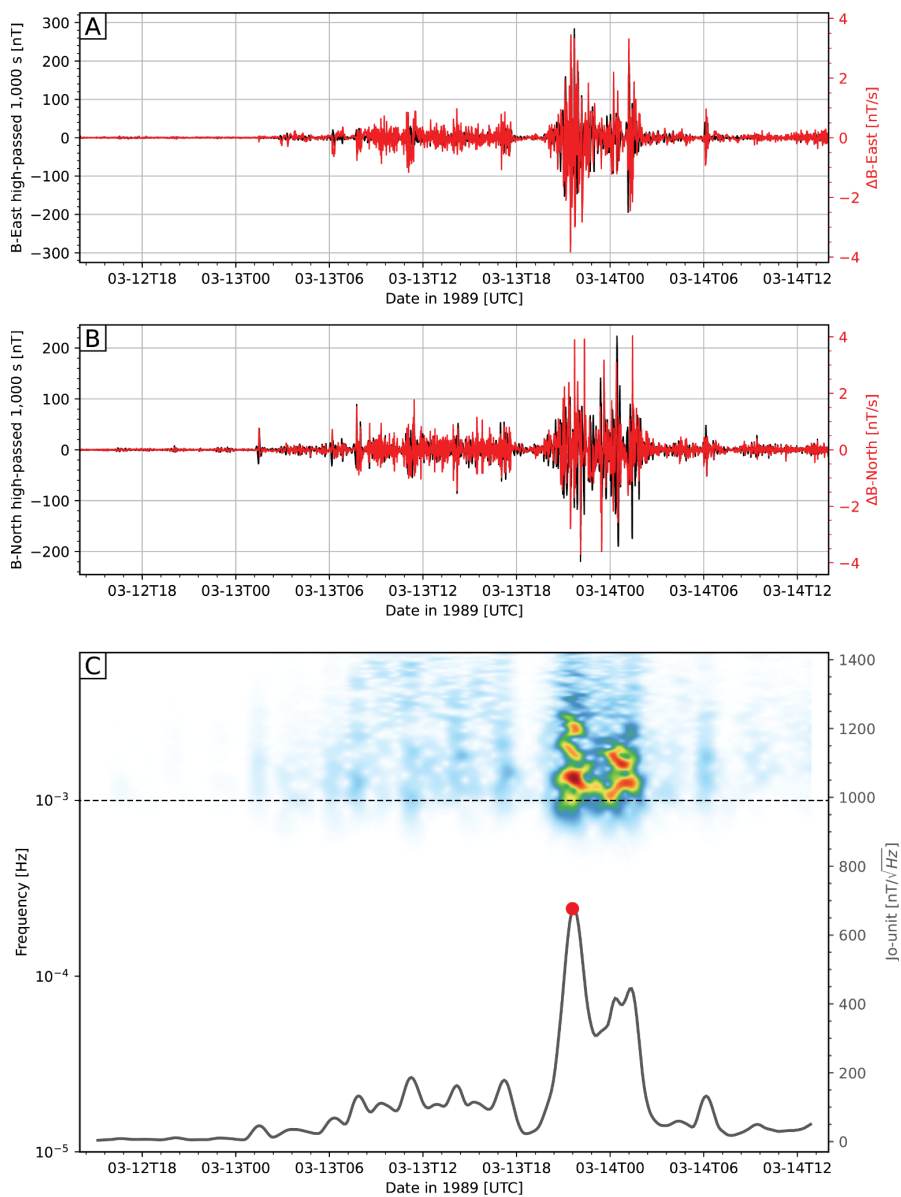

**Figure S3**. Jo-unit measurement at station FRD for the 1989 geomagnetic storm. (A) and (B) show the horizontal components of the geomagnetic field high-pass filtered at 0.001 Hz (black time series), and the time-derivative of the broadband signals (red time series). Note that there is almost negligible difference between the filtered and time-differentiated signals. (C) Shows the average amplitude spectrum of the filtered signals in (A) and (B), and its Jo-unit representation (gray curve). The red circle in (C) marks the peak Jo-unit value for this particular storm.

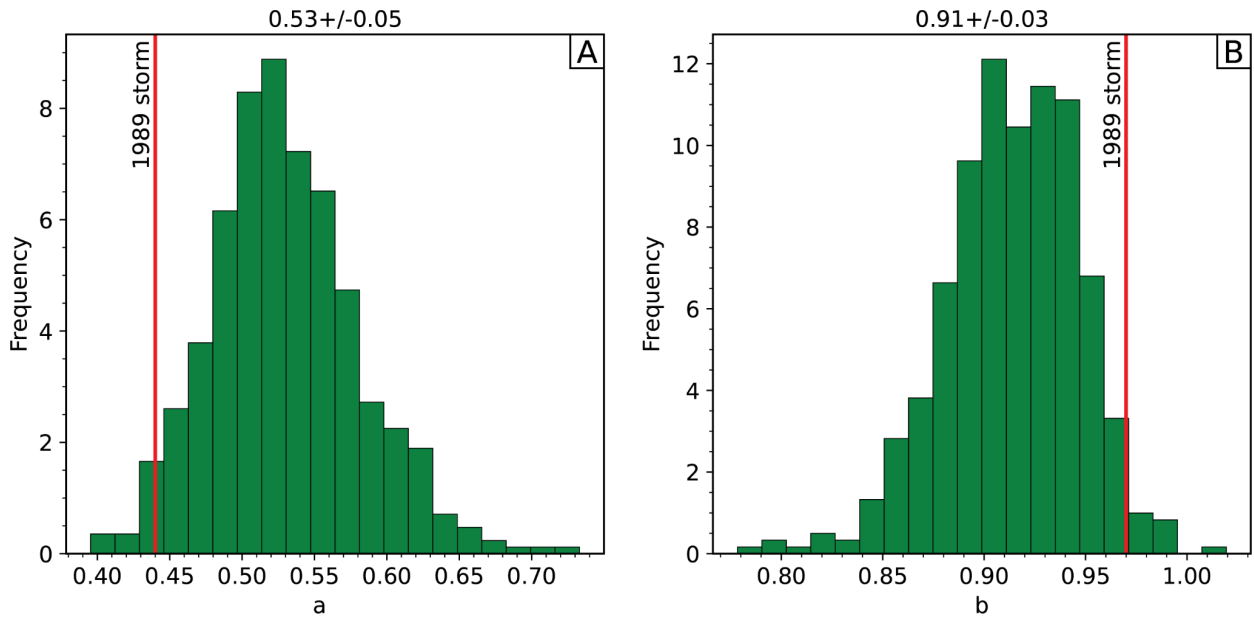

**Figure S4**. Distribution of the bootstrapped parameters of equation 2 for all data points in Figure 5, except for the one for the March 1989 storm. The red line marks the set of parameters that would predict the effects of the 1989 storm on a trans-Atlantic system. The mean and standard deviation of the bootstrapped families is presented on top of each panel.

**Text S2. Attenuation of Electromagnetic waves in marine environments**

It is tempting to assume that the presence of salted seawater makes submarine cables more susceptible to GICs. While it is true that the ocean does increase the overall conductance of the medium, the penetration of the geomagnetic field into the solid Earth also depends on the ground conductivity. This dependence can be easily explained by looking into the governing equation of the intensity of a magnetic field, H,

$$\nabla^2 H \; = \; \mu\sigma\frac{\partial H}{\partial t},$$

where $\mu$ is the magnetic permeability and $\sigma$ is the electrical conductivity. In a half-space, such as the Earth, the frequency domain equivalents of the equation above can be expressed as:

$$\frac{\partial H}{\partial z^2} + k^2 H \; = \; 0,$$

with the solution,

$$H \; = \; H_0 e^{-ikz} = H_0 e^{-i\sigma z} e^{-\beta z},$$

where $H_0$ is the value of the magnetic field at the surface ($z = 0$), $k$ is the propagation constant or complex wave number in the medium, and where,

$$\beta = \sqrt{\frac{\sigma\mu\omega}{2}},$$

with $\omega$ being angular frequency. Thus, the intensity of the magnetic field varies sinusoidally with depth, and experiences a depth-dependent attenuation due to the $e^{-\beta z}$ term in the equation above. A popular way of conceptualizing this phenomena is by looking at the so-called skin depth of an incident magnetic field, which is the depth at which the magnetic field is reduced in amplitude by a factor of $e^{-1}$ or about 37%. Figure S5 shows skin depth for various half-space resistivities. Lower resistivity layers attenuate magnetic fields more rapidly and thus have smaller skin depth than lower conductivity layers.

Seawater has low resistivity (~0.3-Ω/m) and thus a small skin depth. Thus, short-period electromagnetic waves are severely attenuated at seafloor depths, and only long period magnetic fields are able to penetrate the solid Earth (Chave et al., 1999).

Because the geomagnetic field variations occur across periods of about 10 to 1,000 s, we can thus consider the ocean as an insulating layer that prevents GIC in submarine cables.

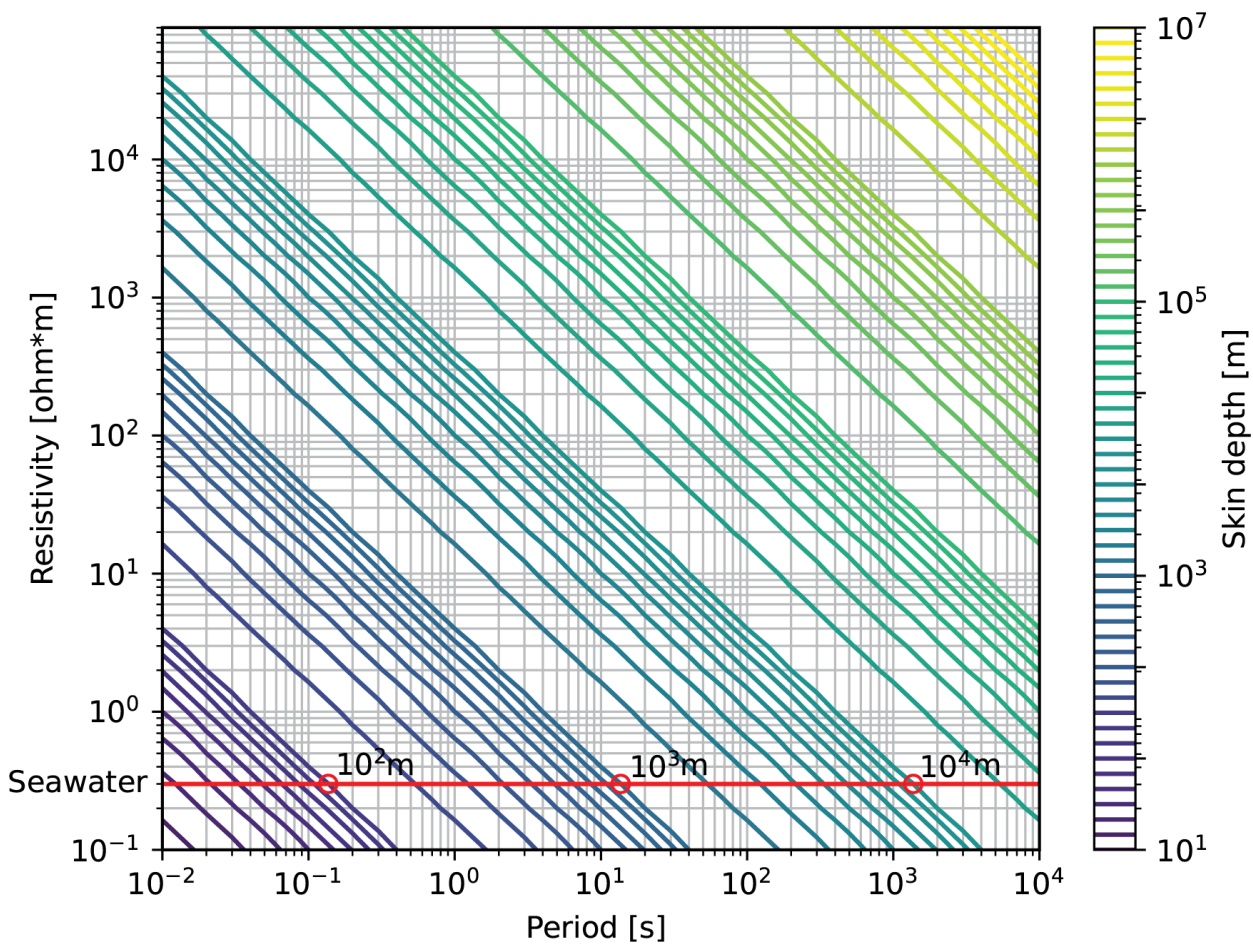

**Figure S5**. Skin depth frequency dependence for various half-space resistivities.

**Table S1.** Length of the systems, date range of the datasets, and name of the magnetic observatories that were used in this investigation.

| System | Length | Measurement time period | Magnetometers |
|---|---|---|---|
| Atlantic E-W #1 | 6,600 km | 2021-11-25 / 2021-12-31 | SBL VAL |
| Atlantic E-W #2 | 5,860 km | 2021-11-25 / 2021-12-31 | FRD CLF |
| Pacific N-S | 10,500 km | 2021-11-01 / 2021-12-01 | FRN PIL |
| Pacific E-W | 9,620 km | 2017-09-01 / 2017-10-01 | KAK FRN |